\documentclass[twocolumn]{aastex631}
\usepackage{amsmath}
\usepackage{hyperref}
\usepackage[nameinlink,capitalise,noabbrev]{cleveref}
\usepackage{soul}
\usepackage[nolist]{acronym} 

\newcommand{\unit}[1]{\ensuremath{\, \mathrm{#1}}}
\graphicspath{{./}{figures/}}

\received{XXX}
\revised{XXX}
\accepted{XXX}

\defcitealias{yun2024}{Paper~I}

\shorttitle{The VSI in Stratified Disks}
\shortauthors{Yun et al.}

\begin{document}
\title{Vertical Shear Instability in Thermally-Stratified Protoplanetary Disks: II. Hydrodynamic Simulations and Observability}

\author[0000-0003-4353-294X]{Han-Gyeol Yun}
\affiliation{Department of Physics \& Astronomy, Seoul National University, Seoul 08826, Korea}
\affiliation{SNU Astronomy Research Center, Seoul National University, 1 Gwanak-ro, Gwanak-gu, Seoul 08826, Republic of Korea}

\author[0000-0003-4625-229X]{Woong-Tae Kim}
\affiliation{Department of Physics \& Astronomy, Seoul National University, Seoul 08826, Korea}
\affiliation{SNU Astronomy Research Center, Seoul National University, 1 Gwanak-ro, Gwanak-gu, Seoul 08826, Republic of Korea}

\author[0000-0001-7258-770X]{Jaehan Bae}
\affiliation{Department of Astronomy, University of Florida, Gainesville, FL 32611, USA}

\author[0000-0002-2641-9964]{Cheongho Han}
\affiliation{Department of Physics, Chungbuk National University, Cheongju 28644, Republic of Korea}

\email{hangyeol@snu.ac.kr, wkim@astro.snu.ac.kr, jbae@ufl.edu, cheongho@astroph.chungbuk.ac.kr}

\begin{acronym}
  \acro{PPD}{protoplanetary disk}
  \acrodefplural{PPD}{protoplanetary disks}
  \acro{3D}{three-dimensional}
  \acro{ALMA}{Atacama Large Millimeter/submillimeter Array}
  \acro{MRI}{magnetorotational instability}
  \acro{MHD}{magnetohydrodynamic}
  \acro{VSI}{vertical shear instability}
  \acro{FWHM}{full-width at half maximum}
\end{acronym}

\begin{abstract}
We conduct three-dimensional hydrodynamic simulations to investigate the nonlinear outcomes and observability of vertical shear instability (VSI) in protoplanetary disks.
Our models include both vertically isothermal and thermally stratified disks, with the latter representing realistic conditions featuring a hotter atmosphere above the midplane. We find that the VSI grows more rapidly and becomes stronger in thermally stratified disks due to enhanced shear, resulting in higher levels of turbulence. At saturation, the turbulence stress reaches $\alpha_{R\phi}\gtrsim 10^{-3}$, more than an order of magnitude stronger than the isothermal case. 
The saturated turbulence is more pronounced near the disk surfaces than at the midplane. On synthetic velocity residual maps, obtained by subtracting the Keplerian rotational velocity, perturbations driven by the VSI manifest as axisymmetric rings in isothermal disks and as ring segments in thermally stratified disks. The latter are visible at disk inclinations as high as $45^\circ$ in thermally stratified disks. The amplitudes of these residual velocities range from $\sim 50$ to $\sim100\unit{m}\;\unit{s}^{-1}$ at a $20^\circ$ inclination, with larger values corresponding to greater thermal stratification. The magnitude of the observed velocity residual increases with the optical depth of the tracer used, as optically thick lines probe the regions near the disk surfaces. 
\end{abstract}


\section{Introduction} \label{sec:intro}

Gas in a \ac{PPD} gradually moves radially inward while following nearly Keplerian rotation around a central star. \citet{ss1973} and \citet{lp1974} attributed mass accretion to turbulent viscosity, which transports angular momentum radially outward. Since then, numerous studies have investigated the mechanisms that generate turbulence within \acp{PPD} (see \citealp{le2023} and references therein). In hot accretion disks surrounding compact objects, the \ac{MRI} identified by \citet{bh1991} has long been considered the most promising source of turbulence. Due to self-shielding, however, the majority of gas in \acp{PPD} has a low ionization fraction and temperature, which suppresses \ac{MRI} activity except in the surface layers. Various non-ideal \ac{MHD} simulations have supported the notion of inefficient \ac{MRI} in driving turbulence in \acp{PPD} \citep{bs2013, sb2013, lk2014}. This raises the question of what mechanisms drive turbulence and mass accretion in these disks.

The \ac{VSI} is a potential mechanism for turbulence driving within a \ac{PPD}. Its operation requires a vertical variation of rotational angular velocity, which is readily achieved in \acp{PPD} due to the decreasing gravitational influence of the central star with increasing height and the decreasing disk temperature with radius \citep{ub1998, ng2013}. Although originally proposed for angular momentum transport in the radiative zone of a differentially rotating star \citep{gs1967, f1968}, the \ac{VSI} has gained attention in the context of accretion disks following various analytical \citep{ub1998, u2003} and numerical studies \citep{ng2013, sk2014}. Since the \ac{VSI} is purely a hydrodynamical process, it can operate even in cold regions where \ac{MRI} is suppressed.

Previous numerical studies have revealed various characteristics of the \ac{VSI}, including the meridional circulation pattern of its perturbation, which exhibit a radially narrow and vertically elongated structure \citep{ng2013, sk2014}. This results in anisotropic turbulence with stronger vertical motion compared to radial motion \citep{sk2017}. Additionally, it has been shown that the \ac{VSI} can persist despite stabilization by vertical buoyancy and magnetic fields \citep{ng2013, lp2018}, as well as through radiative diffusion, \citep{sk2014} and non-ideal \ac{MHD} effects \citep{cb2020}.

While the theoretical and numerical studies mentioned above suggest the \acp{PPD} are likely turbulent presumably due to the \ac{VSI}, observational confirmations have been lacking. Recent advancements in spectral and spatial resolution, as well as sensitivity, have made it possible to measure turbulence levels in \acp{PPD} (\citealp{fh2015, tg2016}; see reviews by \citealp{r2023}). Direct measurements of molecular line broadenings in sub-millimeter wavelengths suggest that turbulence in \acp{PPD} is at the level of $\delta v_{\mathrm{turb}} \lesssim 10^{-1}c_s $, with $c_s$ being the isothermal speed of sound, in disks around HD163296, TW Hya, MWC 380 and V4046 Sgr \citep{fh2015, fh2017, fh2018, fh2020, th2018}. Indirect measurements based on dust diffusion \citep{db2018, fb2020, rt2020}, disk size \citep{tr2020}, and accretion rate \citep{rb2019} also indicate similar upper limits for turbulence levels in the disks. These levels are overall comparable to $\delta v_{\mathrm{turb}} \sim 0.05c_s$ predicted by numerical simulations of \ac{VSI} in \acp{PPD} \citep{fn2017, ft2020}.

Recent observational techniques also now enable the probing of the dynamical structure of disks and the detection of small velocity perturbations on the order of $\sim 10\unit{m\ s^{-1}}$ \citep{tb2021}. This advancement opens up the possibility of detecting anisotropic turbulent motions in PPDs using molecular lines with \ac{ALMA}. \citet{bf2021} recently reported that turbulent motions induced by the \ac{VSI} are significant enough to be detectable with \ac{ALMA}, appearing as various quasi-axisymmetric rings in velocity residual maps derived from molecular line emission images.

While the numerical models of \citet{bf2021} suggest that \ac{VSI}-driven turbulence may possess observable signatures, they were restricted to vertically isothermal disks.
However, real disks naturally exhibit thermal stratification caused by irradiation from the central star, which heats the surfaces preferentially, leaving optically thick regions near the midplane \citep[see, e.g.,][]{cg1997, dc1998, dc1999}.
As the vertical profile of the rotational velocity is influenced by the pressure gradient, the characteristics of the \ac{VSI} in thermally stratified disks are anticipated to differ from those in isothermal counterparts.

In our previous work, \citet[][hereafter Paper I]{yun2024}, we examined the linear stability of the \ac{VSI} in thermally stratified disks using both local and semi-global approaches. Our analysis revealed that the \ac{VSI} manifests in two distinct modes: surface and body modes \citep[see also][]{ng2013,bl2015}. Surface modes localize in regions with strong shear, while body modes are distributed throughout the main body of a disk. In thermally stratified disks, surface modes bifurcate into two branches. The branch associated with the strongest shear at mid-height regions exhibits higher growth rates compared to the other branch occurring at the disk surfaces. Generally, surface modes, which are present only for sufficiently large radial wavenumbers, display higher growth rates than body modes. Furthermore, thermal stratification of the disk enhances the growth rates of both surface and body modes due to increased vertical shear, and amplifies the \ac{VSI}-driven kinetic energy in the radial direction relative to the vertical direction.

This paper follows up on \citetalias{yun2024} by conducting hydrodynamic simulations of thermally stratified disks susceptible to the \ac{VSI}. This work is a straightforward extension of the isothermal models of \citet{bf2021} to disks with a vertical temperature gradient. We investigate the nonlinear consequences of the \ac{VSI} in these stratified disks and assess the observability of a saturated turbulent state. We also explore how the optical depth of tracer molecules affects velocity residual maps in thermally stratified disks.

This paper is organized as follows: 
In \autoref{sec:methods}, we describe our \ac{3D} disk models and numerical methods. 
In \autoref{sec:results}, we present the results of our nonlinear simulations and the synthetic velocity residual maps at turbulence saturation. 
In \autoref{sec:discussion}, we discuss the contribution of each velocity component to the velocity residual maps and the effects of the optical depth of tracer molecules. Finally, we summarize our results in \autoref{sec:conclusion}.

\section{Numerical Methods} \label{sec:methods}

We adopt the same disk models as in \citetalias{yun2024}. We set the disk temperature as 
\begin{equation}
  T(R,Z)=
  \begin{cases}
  \displaystyle T_{\text{atm}}(R) + \left[T_{\text{mid}}(R)-T_{\text{atm}}(R)\right]\cos^2\left(\frac{\pi Z}{2Z_q }\right), \\
  \hfill \text{for $Z < Z_q$}, \\
  T_{\text{atm}}(R), \hfill \text{for $Z \geq Z_q$},
  \end{cases}
  \label{eq:temp}
\end{equation}
where $T_{\text{mid}}(R)=25.7(R/R_0)^{-1/2}\unit{K}$ is the temperature in the midplane,
with $R_0=100\unit{au}$ being the reference radius, and $T_\text{atm}$ is the temperature in the disk “atmosphere” located at $Z \geq Z_q$ \citep{dd2003}. We set the pressure scale height to $H =10(R/R_0)^{5/4}\unit{au}$, and take $Z_q=3H$. We consider three models with $n\equiv T_\text{atm}/T_\text{mid}=1,2,3$, with $n=1$ corresponding to the vertically isothermal disk. 

Our disks are initially in hydrostatic equilibrium along the vertical direction, with the density distribution given by 
\begin{align}
  \rho(R,Z) =
  &\rho_{0} \left(\frac{R}{R_0}\right)^{-9/4} \frac{c_s^2(R,Z)}{c_s^2(R,0)}\times \notag \\
  &\exp\left[-\int_0^{Z}\frac{1}{c_s^2(R,Z')}\frac{\partial\Phi}{\partial Z'}dZ'\right],
\end{align} 
where $\rho_{0}$ is the density at $(R,Z)=(R_0,0)$, $c_s=(k_BT/\mu m_\text{H})^{1/2}$ is the isothermal speed of sound with the mean molecular weight $\mu=2.3$, and $\Phi(R,Z) = -GM_*/(R^2+Z^2)^{1/2}$ is the gravitational potential of the central star with mass $M_*=1\unit{M_\odot}$. 
We take $\rho_{0}\approx 0.018M_* R_0^{-3}$, for which the total disk is 5\% of the mass of the central star. 

The radial force balance requires the angular velocity $\Omega$ of the disk rotation satisfies 
\begin{equation}\label{eq:eqOmg}
  \Omega^2(R,Z) = \Omega_K^2\sin\theta + \frac{1}{\rho r\sin^2\theta}\frac{\partial (c_s^2\rho)}{\partial r},
\end{equation}
where $\Omega_K(R) = \sqrt{GM_*/R^3}$ is the Keplerian angular velocity, $\theta=\cos^{-1}(Z/r)$ is the polar angle, and $r=R/\sin\theta$ is the spherical radius. The vertical shear can be measured by the shear parameter defined as 
\begin{equation}
q = - R\frac{\partial \ln \Omega}{\partial Z}. 
\end{equation}
As illustrated in Figure 1($d$) of \citetalias{yun2024}, a disk with a larger $n$ exhibits higher vertical shear. In the isothermal disk with $n=1$, $q$ increases approximately linearly with $Z$. In contrast, in the stratified disks with $n=2$ and 3, $q$ takes on a parabolic shape at $Z<Z_q$ where $\partial T/\partial z$ is positive, peaking at $Z\sim 0.5Z_q$, and a linear trend at $Z\geq Z_q$ where $T$ is constant. The density-weighted vertically averaged shear parameter is $\langle q\rangle = 0.02$, 0.06, and 0.10 for the disk with $n = 1$, 2, and 3, respectively, suggesting that the \ac{VSI} should be stronger in a disk with larger $n$. 

We use FARGO3D \citep{bm2016} to run \ac{3D} hydrodynamic simulations in spherical polar coordinate $(r,\theta,\phi)$ by adopting a locally isothermal equation of state. Our computational domain is equal to that of \citet{bf2021}, which
spans $r$ from $40\unit{au}$ to $250\unit{au}$ in the radial direction and $\theta$ ranging from $\pi/2-\arctan(4.5H_0/R_0)$ to $\pi/2+ \arctan(4.5H_0/R_0)$ in the meridional direction, while covering the entire $2\pi$ radians in the azimuthal direction. Here, $H_0=H(R_0)=10\unit{au}$. 
The domain is discretized into $(512,128,1024)$ cells in the 
$(r,\theta,\phi)$ directions, respectively. We employ a logarithmic grid spacing in the radial direction with $\Delta r = 0.35\unit{au}$ at $R=R_0$, while using linear spacing in both meridional and azimuthal directions with $\Delta\theta = H_0/(15.1 R_0)$ and $\Delta\phi=6.1\times10^{-2}\unit{radians}$. To seed perturbations for the \ac{VSI}, we initially introduce random noises to the meridional velocity $v_\theta$ at the level of $\sim 10^{-6}c_s$. 

At the meridional boundaries, we implement the zero-gradient boundary conditions for $v_r$ and the reflecting boundary conditions for $v_{\theta}$, while extrapolating $\rho$ and $v_{\phi}$ on ghost cells from adjacent active cells. 
At the radial boundaries, we apply the reflecting boundary conditions for $v_r$ and the zero-gradient boundary conditions for $v_{\theta}$, and extrapolate $\rho$ and $v_{\phi}$. In addition, we establish wave-damping zones at $40\unit{au}\leq r \leq 48\unit{au}$ and $200\unit{au}\leq r \leq 250\unit{au}$ to prevent wave reflection \citep{de2006}.
To ensure numerical convergence, we conduct additional simulations with halved grid spacing and various meridional boundary conditions. The results presented in \autoref{sec:test} demonstrate that the numerical outcomes remain robust despite changes in resolution and boundary conditions.

\section{Results} \label{sec:results}

In this section, we first describe the simulation results, focusing on the morphologies and the level of turbulence at saturation. We then present the synthetic observations of the model disks using various CO lines.

\begin{figure}
  \epsscale{1.15}
  \plotone{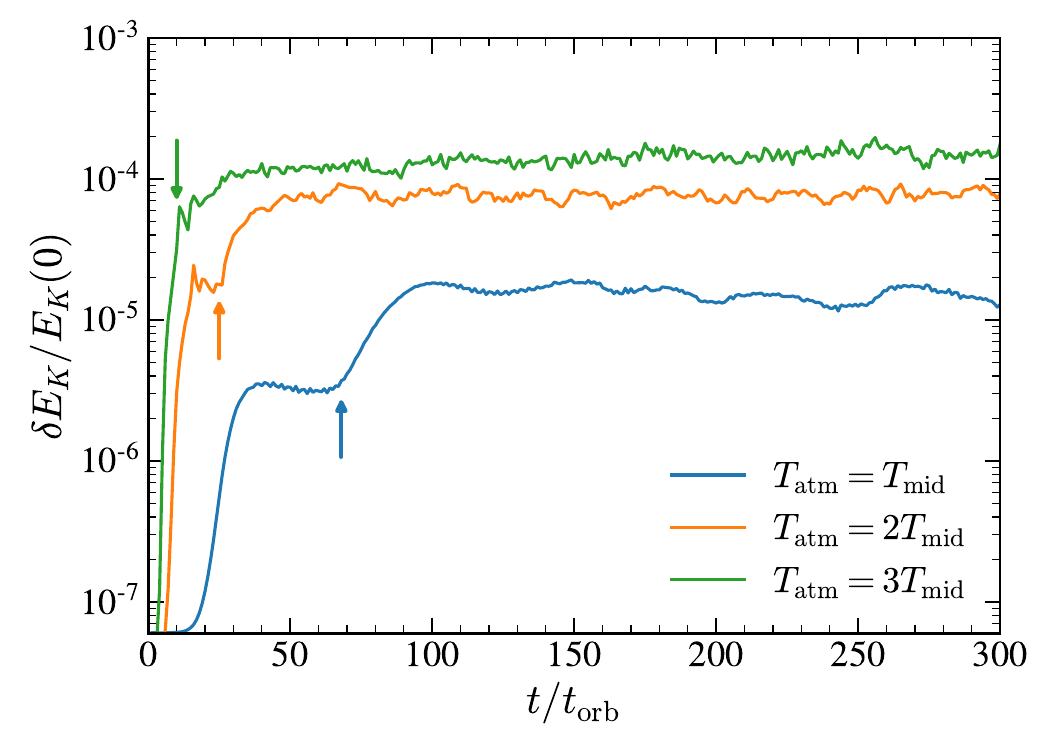}
  \caption{
  Temporal variations of the perturbed kinetic energy $\delta E_K$ normalized by the initial kinetic energy $E_K(0)$ for models with $T_\textrm{atm}/T_\textrm{mid}=0,1,2$. The initial increase in $\delta E_K$ results from the growth of surface modes, while the subsequent growth phase, indicated by an arrow in each model, is driven by body modes.}
  \label{fig:Ek_total}
\end{figure}

\begin{figure*}
  \epsscale{1.1}
  \plotone{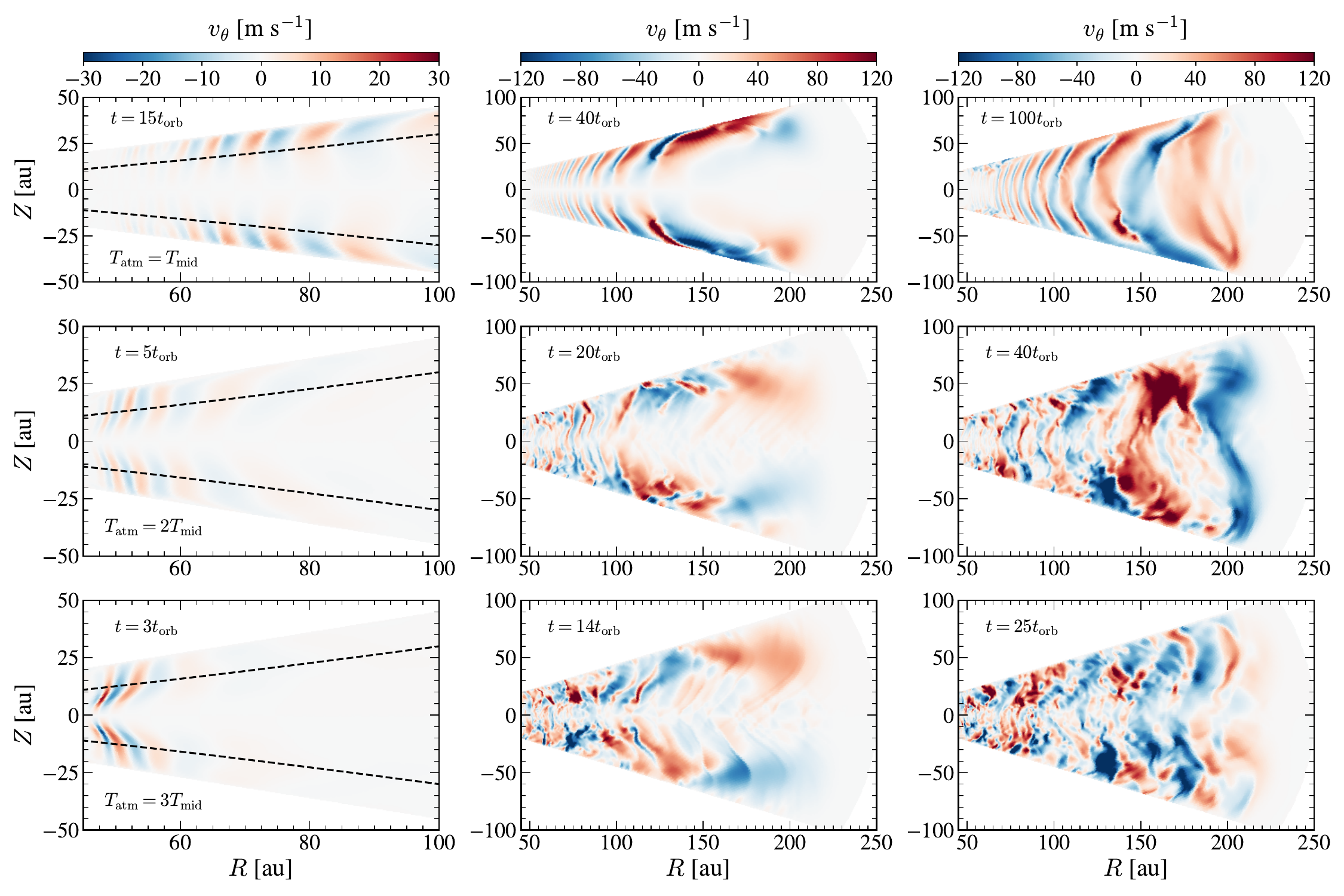}
  \caption{
  Snapshots of the meridional velocity $v_{\theta}$ in the $R$--$Z$ plane at selected times. Each row corresponds to the models with $n=1$, 2, and $3$ from top to bottom, respectively. The dashed lines in the first column marks the height with $|Z|=Z_q$. In all models, \ac{VSI} perturbations grow fastest near the disk atmospheres at smaller radii in the early stages. The \ac{VSI}-unstable modes have higher growth rates and longer radial wavelengths in disks with stronger thermal stratification.}
  \label{fig:evol}
\end{figure*}

\begin{figure}
\plotone{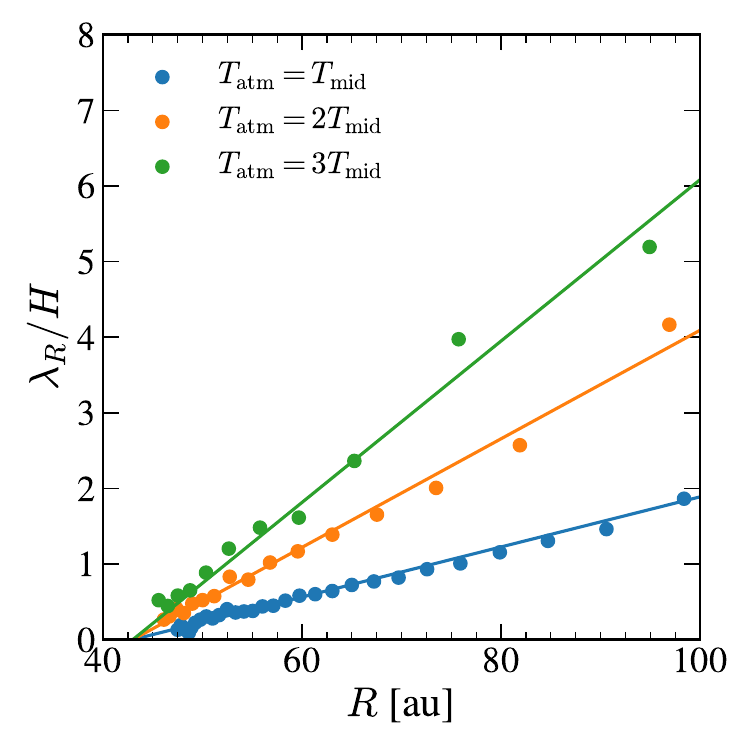}
  \caption{Radial wavelengths $\lambda_R$ of the VSI measured at the times shown in the left panels of \cref{fig:evol}. The solid lines are the best fits described by \cref{eq:fit}, indicating that $\lambda_R$ is longer at larger $R$ and in a more thermally stratified disk.
  }
  \label{fig:wv2}
\end{figure}

\subsection{Hydrodynamic Simulations}\label{sec:HDsim}

\begin{figure*}
  \epsscale{1.1}
  \plotone{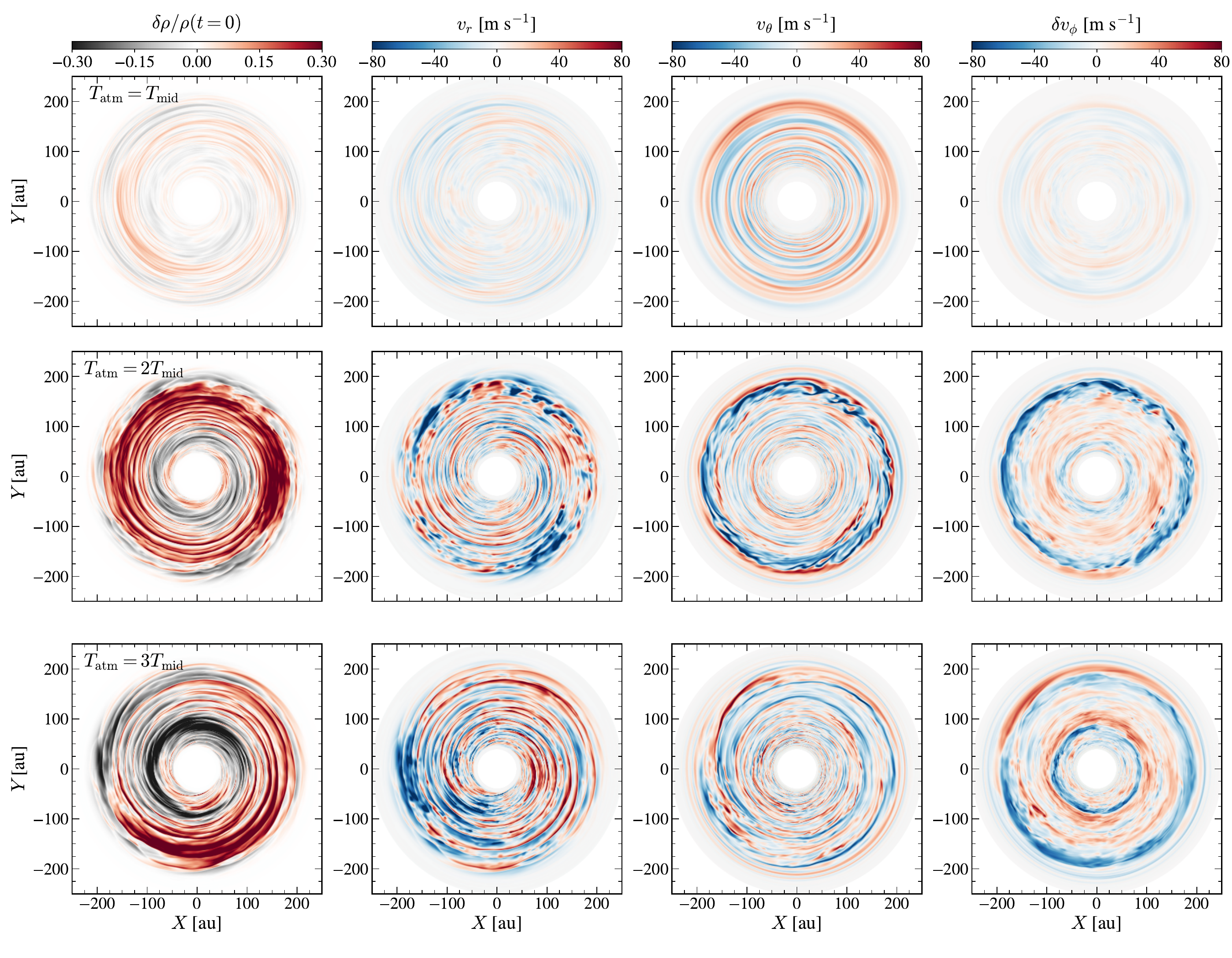}
  \caption{
  Distributions of the perturbed quantities at the midplane at $t/t_{\text{orb}}=300$. Each column represents the perturbed density $\delta \rho \equiv \rho-\rho(t=0)$, radial velocity $v_r$, meridional velocity $v_{\theta}$, and perturbed azimuthal velocity $\delta v_\phi \equiv v_{\phi}-v_{\phi}(t=0)$ from left to right, respectively. Each row corresponds to the model with $n=1$, 2, and 3 from top to bottom, respectively. The \ac{VSI}-induced perturbations are stronger in disks with a higher degree of thermal stratification. While the velocity perturbations are predominantly meridional in the isothermal disk, the thermally stratified disk exhibits significant perturbations in the radial and azimuthal velocities as well.}
  \label{fig:per1}
\end{figure*}

\begin{figure*}
  \epsscale{1.1}
  \plotone{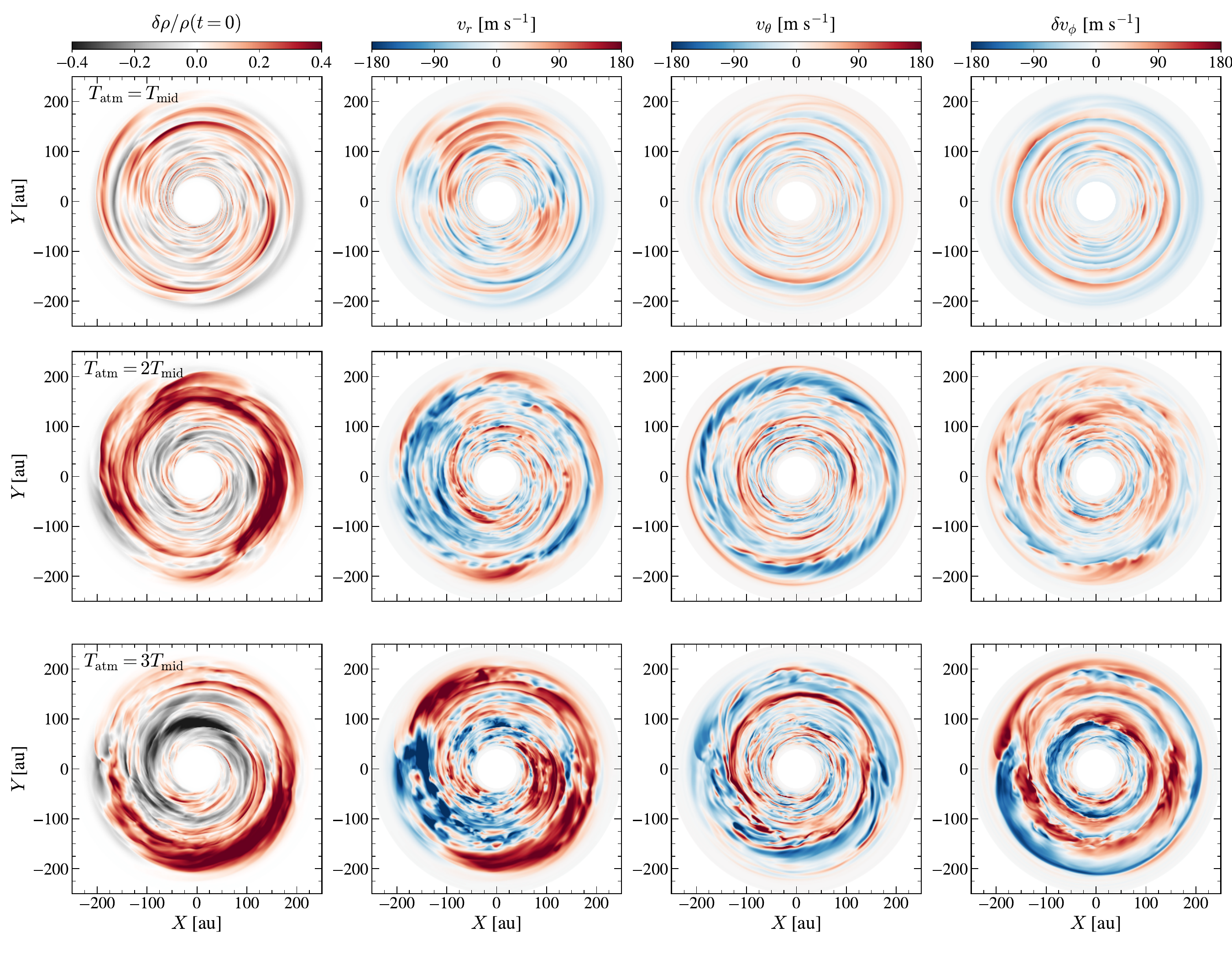}
  \caption{Same as \cref{fig:per1} but depicted at the high-altitude region with $z=3H$.}
  \label{fig:per2}
\end{figure*}

\begin{figure*}
  \epsscale{1.1}
  \plotone{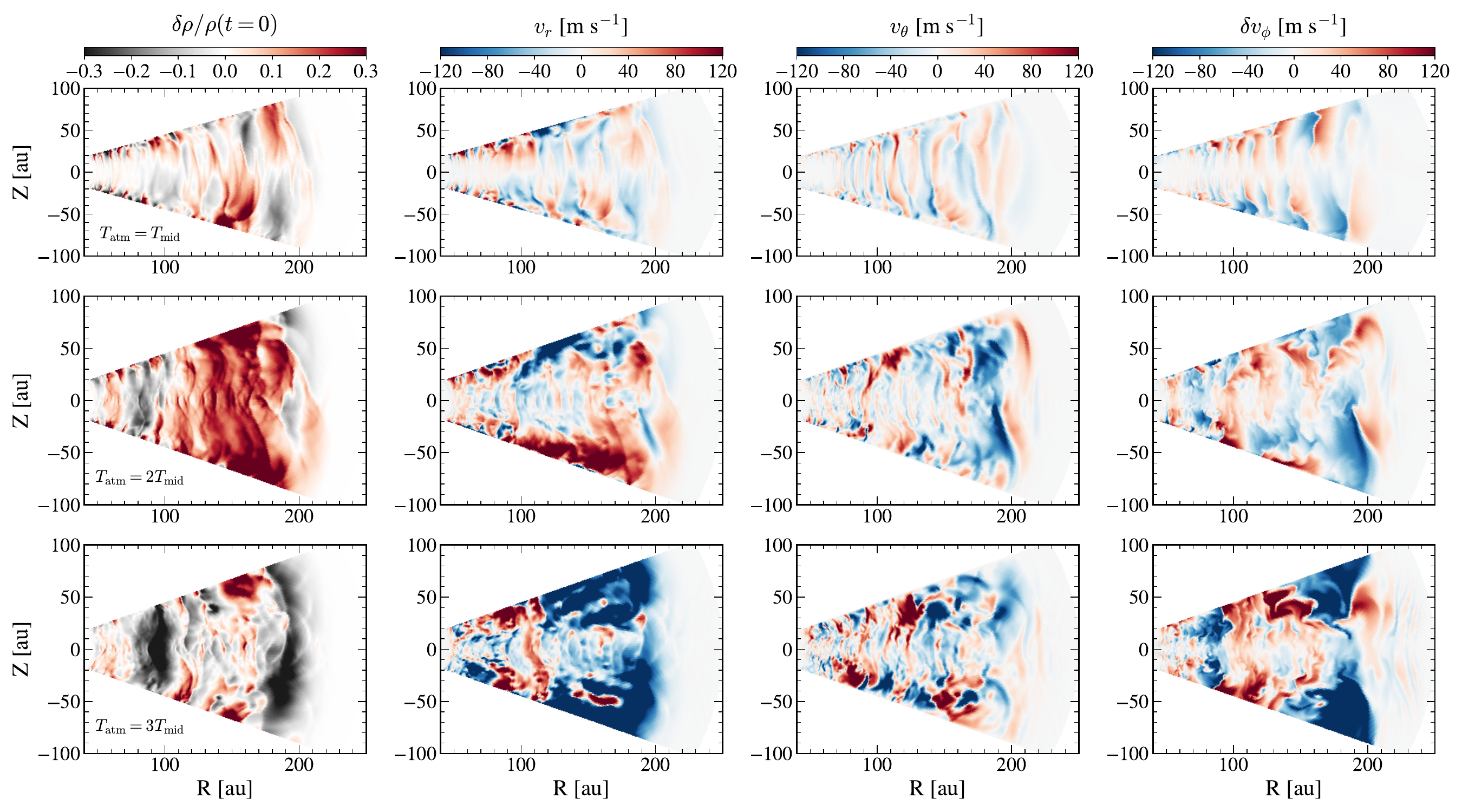}
  \caption{
  Distributions of the perturbations in the $R$--$Z$ plane at $\phi = 0$ when $t/t_{\text{orb}}=300$. The panel layout is the same as in \cref{fig:per1}. The turbulence driven by the \ac{VSI} is generally stronger at the surface compared to the midplane and also in a model with a larger degree of thermal stratification.
  }
  \label{fig:per3}
\end{figure*}

\Cref{fig:Ek_total} plots the evolutionary histories of the perturbed kinetic energy $\delta E_K\equiv \case{1}{2} \int \rho (v_r^2 + v_\theta^2 + \delta v_\phi^2) dV$ relative to the initial kinetic energy $E_K(0)\equiv \case{1}{2} \int \rho  v_\phi^2 dV$ at $t=0$ for all models as functions of $t/t_\mathrm{orb}$. 
Here, $\delta v_\phi\equiv v_\phi - v_\phi(t=0)$ is the perturbed azimuthal velocity and $t_\mathrm{orb}=2\pi/\Omega(R_0)$ is the orbital time at the reference radius. 
In each disk, the initial perturbations applied to $v_\theta$ induce wave motions in the gas. The system undergoes an initial relaxation phase as the waves propagate throughout the disk. Once the system picks up the most unstable \ac{VSI} modes, $\delta E_K$ begins to grow with time. This occurs at $t/t_\text{orb}\sim 15$ in the isothermal disk with $n=1$, but much earlier at $t/t_\text{orb}\lesssim 3$ in the thermally stratified disks with $n=2$ and $3$.

According to \citet{ng2013}, the \ac{VSI} comprises two distinct modes: surface modes (or finger modes) which develop near the disk surfaces and body modes growing near the midplane. These surface modes typically have shorter radial wavelengths and undergo faster growth, while the body modes have longer radial wavelengths and slower growth rates (\citetalias{yun2024}). In \Cref{fig:Ek_total},
$\delta E_K$ exhibits two discernible phases of exponential growth. In the isothermal disk with $n=1$, $\delta E_K$ exhibits an initial exponential growth (at $15\lesssim t/t_\textrm{orb}\lesssim 30$) and saturation
(at $t/t_\textrm{orb}\sim 30$) of the surface modes. The perturbations induced by the surface modes propagate toward the midplane, thereby exciting body modes. The slower-growing body modes begin to dominate at $t/t_\textrm{orb}\sim 70$, indicated by the blue arrow in \Cref{fig:Ek_total}, causing a further increase in $\delta E_K$ around $t/t_\textrm{orb}\sim70$--100, ultimately saturating at $\delta E_K/E_K(0)\sim10^{-5}$. As the disk becomes thermally stratified, both surface and body modes grow more rapidly, and the interval between the first and second growth phases of $\delta E_K$ shortens. In the $n=2$ disk, the saturation of the surface modes occurs at $t/t_\textrm{orb}\sim 20$,  with body modes beginning to dominate at $t/t_\textrm{orb}\sim 25$. In contrast, in the $n=3$ disk, the body modes dominate as early as $t/t_\textrm{orb}\sim 10$, even before the surface modes saturate.
The saturated level of $\delta E_K$ is also higher in a more thermally stratified disk. 

To analyze the impact of thermal stratification on the \ac{VSI} in detail, we examine the induced velocity perturbations. \cref{fig:evol} presents snapshots of the meridional velocity $v_\theta$ in the $R$--$Z$ plane at selected epochs. Two notable differences emerge between the isothermal and thermally stratified disks. Firstly, the growth of the \ac{VSI} is relatively slow in the isothermal disk. However, in thermally stratified disks, the initial relaxation phase is brief, and the \ac{VSI} grows more rapidly, reaching turbulence saturation sooner. Secondly, while the \ac{VSI} initially arises near the disk surfaces in the isothermal disk, in stratified disks, it emerges in regions with $|Z| \approx 0.5 Z_q$, where the vertical shear is the strongest. This is most clearly seen in the $n=3$ model. Over time, the regions experiencing significant growth of the \ac{VSI} extend radially outward, leading to an increase in the perturbed velocities. Concurrently, the \ac{VSI} modes propagate from the surfaces to the midplane, engaging in nonlinear interactions with modes propagating from opposite sides and different radii. These nonlinear interactions among \ac{VSI} modes drive the disk into a highly turbulent state. 

As shown in the left panels of \cref{fig:evol}, 
the radial wavelength of the dominant \ac{VSI} mode in a thermally stratified disk is longer than that in the vertically isothermal disk. To estimate the most susceptible wavelength of the VSI during the linear phase in the simulations, we take the data for the vertical momentum $\rho v_z$ at the times shown in the left panels of \autoref{fig:evol}, which allows to capture the initial growing phase of the
\ac{VSI}. We measure the radial distances between two consecutive zeros of $\rho v_z$ at $Z/H=Z_q/2$ and set it equal to half of the radial wavelength $\lambda_R$ \citep[see, e.g., ][]{sc2022}. 
\Cref{fig:wv2} plots the radial distributions of $\lambda_R/H$ as functions of $R$. Our best fits are 
\begin{equation}\label{eq:fit}
\frac{\lambda_R}{H}=\begin{cases} 3.31(R/R_0-0.43), & \text{for $n=1$},\\
7.18(R/R_0-0.43), & \text{for $n=2$},\\
10.7(R/R_0-0.43), & \text{for $n=3$}.
\end{cases}
\end{equation}
Note that the radial wavelength goes to zero at $R/R_0=0.43$ due to the wave-damping zone applied at the inner radial boundary. A more thermally stratified disk has longer $\lambda_R$ due to stronger shear, which is consistent with Equation (10) of \citetalias{yun2024} for fixed $k_zH\sim1$. The increasing trend of $\lambda_R$ with $R$ in the linear growth phase is also similar to the numerical results of \citet{sc2022} in the nonlinear regime.

\Cref{fig:per1,fig:per2} present the face-on distributions of the density and velocity at $z/H=0$ and 3, respectively, while \Cref{fig:per3} plots the edge-on distributions of the azimuthally-averaged quantities, all at $t/t_{\text{orb}}=300$. 
Each column shows the perturbed density $\delta \rho$ and radial velocity $v_r$, meridional velocity $v_\theta$, and perturbed azimuthal velocity $\delta v_\phi$ from left to right. Each row corresponds to the model with $n=1$, 2, and 3 from top to bottom. Clearly, the perturbations induced by the \ac{VSI} are more pronounced at higher $|z|$ and in disks with greater thermal stratification. Additionally, structures of the VSI-induced turbulence in the stratified disks are more complex compared to those in the vertically isothermal disk.

As \Cref{fig:per1,fig:per2} show, the meridional and azimuthal components of the perturbed velocities exhibit quasi-axisymmetric ring-like structures, reflecting the axisymmetric nature of the \ac{VSI} (\citealt{ng2013, sk2014}; \citetalias{yun2024}). This in turn implies that the perturbed azimuthal velocity is smaller than the other components. Since the \ac{VSI} is incompressible and has $|v_r/v_\theta|\sim |k_z/k_x| \sim q$ \citepalias{yun2024}, the density and radial velocity perturbations are weaker than the other perturbations. 
In the isothermal disk, the maximum amplitudes of the  time-averaged velocity perturbations after saturation in the midplane are 21, 57, and 15$\unit{m\ s^{-1}}$ in the radial, meridional, azimuthal directions, respectively. These increase to 82, 95, and 50$\unit{m\ s^{-1}}$ in the $n=2$ disk and 123, 116, and 73$\unit{m\ s^{-1}}$ in the $n=3$ disk.
The corresponding ratio of the perturbed velocities is $|v_r/v_\theta|\sim 0.4$, $0.9$, and $1.1$ and $|\delta v_\phi/v_\theta|\sim 0.3$, $0.5$, and $0.6$ in the disk with $n=1$, 2, and 3, respectively. Although the trend of $|v_r/v_\theta|$ increasing with $n$ aligns with the finding of \citetalias{yun2024}, the velocity ratio is quantitatively larger than the linear prediction, likely due to the combined effect of gas compressibility, nonlinear wave interactions, and other factors. Additionally, it is possible that the VSI in thermally stratified disks grows sufficiently to excite the parasitic instabilities of the Kelvin–Helmholtz type, resulting in the formation of vortices within the disk plane \citep{lp2018}. 

\citet{bf2021} found that the presence of the quasi-axisymmetric rings in the distributions of $v_\theta$ is crucial for detecting the \ac{VSI}-induced turbulence. We note that thermal stratification not only enhances the level of turbulence but also reduces the number of ring-like features. This reduction is due to the fact that strong vertical shear favors long radial wavelengths for growth (Equation  10 of \citetalias{yun2024}). Also, the non-monotonic $q$ profile in the $n=2,3$ disks makes the radial and vertical wavelengths of the VSI modes vary throughout the disk, making it harder for coherent structures to develop.

\begin{figure}
  \epsscale{1.1}
  \plotone{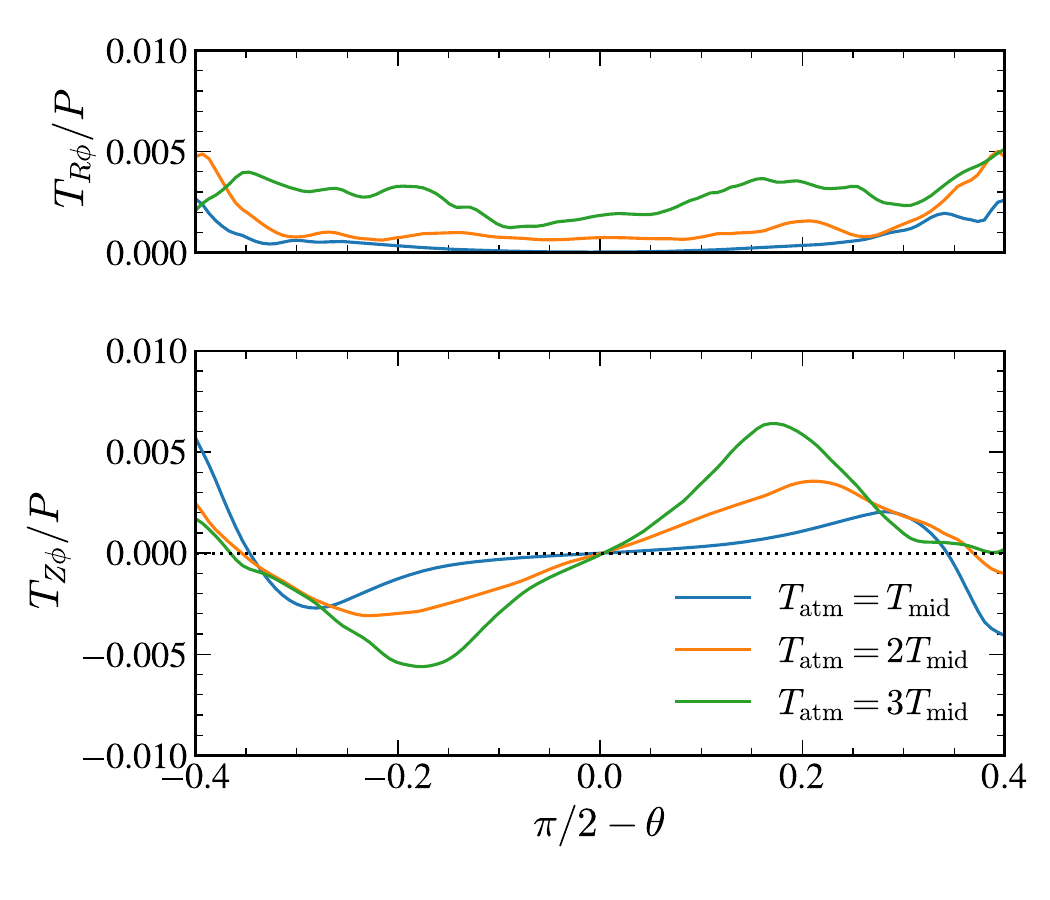}
  \caption{Vertical distributions of the $R$--$\phi$ and $Z$--$\phi$ Reynolds stresses, normalized by the thermal pressure at $r=100\unit{au}$. Thermal stratification not only increases $T_{R\phi}$ and $T_{Z\phi}$ but also shifts the regions of maximum $T_{Z\phi}$ toward $|Z|\sim Z_q/2$, where shear is strongest.}
  \label{fig:rey2d}
\end{figure}

To find out the height where the \ac{VSI} is strongest, we plot in \cref{fig:rey2d} the Reynolds stresses $T_{R\phi}\equiv \langle\rho v_R v_\phi\rangle-\langle v_{\phi}\rangle\langle \rho v_R\rangle$ and $T_{Z\phi}\equiv \langle\rho v_Z v_\phi\rangle-\langle v_{\phi}\rangle\langle \rho v_Z\rangle$, normalized by the thermal pressure $P=\rho c_s^2$ at $r=100\unit{au}$, as functions of the polar angle. Here, $\langle f \rangle$ denotes the average of the quantity $f$ over $r=80$--$120\unit{au}$, $\phi=0$--$2\pi$, and $t=150$--$300\,t_{\rm orb}$. Clearly, $T_{R\phi}$ is more or less symmetric with respect to the midplane, while $T_{Z\phi}$ is antisymmetric \citep[e.g.,][]{zz2024}. Thermal stratification increases the level of both $T_{R\phi}$ and $T_{Z\phi}$. In particular, a higher value of $T_{R\phi}$ at the disk midplane suggests an enhanced radial gas inflow in a more thermally stratified disk, consistent with \citet{zz2024}. In addition, thermal stratification shifts the regions of maximum $T_{Z\phi}$ toward $|Z|\sim Z_q/2$, where shear is strongest.

\begin{figure}
  \epsscale{1.1}
  \plotone{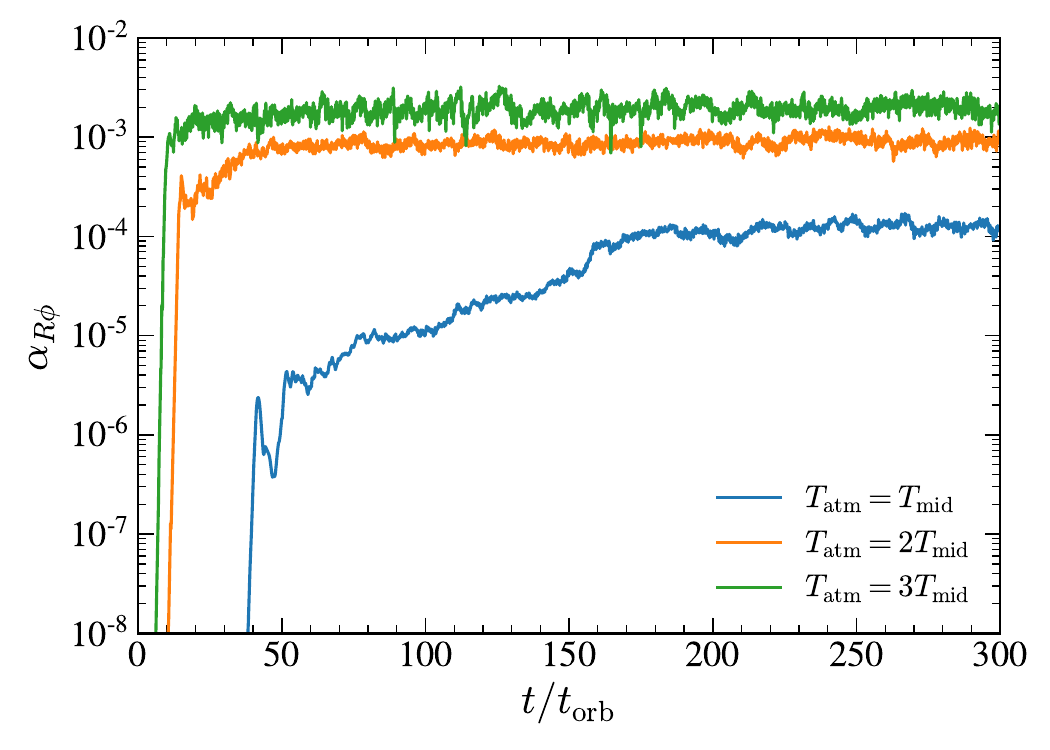}
  \caption{Temporal evolution of the Reynolds stress-to-pressure ratio, $\alpha_{R\phi}$. At the end of the simulation, the thermal stratification increases $\alpha_{R\phi}$ by a factor of $\sim 10$ and $\sim20$ in the disk with $n=2$ and 3, respectively, compared to the isothermal disk.}
  \label{fig:rey}
\end{figure}

To quantify the level of turbulence driven by the \ac{VSI}, we calculate the ratio of the volume-averaged Reynolds stress, $T_{R\phi}$, to the volume-averaged thermal pressure as 
\begin{equation}\label{eq:alpha}
  \alpha_{R\phi} \equiv \frac{\int T_{R\phi} dV}{\int PdV},
\end{equation}
\Cref{fig:rey} plots temporal changes of $\alpha_{R\phi}$ for our models.
In the isothermal disk, $\alpha_{R\phi}$ exhibits a dramatic increase until $t/t_{\mathrm{orb}}\sim60$ when the growth of the surface modes saturates. Subsequently, it grows more slowly, as the body modes develop, increasing the turbulence level. It eventually saturates at $t/t_\textrm{orb}\sim 200$ when the turbulence is fully developed in the disk, reaching a level of $\alpha_{R\phi}\sim 1\times10^{-4}$, consistent with the results of \citet{bf2021}. 
In the thermally stratified disks, the \ac{VSI} begins to grow earlier and faster than in the isothermal disk. The phase of slowly increasing $\alpha_{R\phi}$ is brief, as well. The saturated level of \ac{VSI}-driven turbulence is $\alpha_{R\phi}\sim 1\times 10^{-3}$ and $2\times 10^{-3}$ for the disk with $n=2$ and 3, respectively. 

\subsection{Synthethic Observations}\label{sec:SynObs}

We turn our attention to examining the the observability of \acp{PPD} with turbulent motions driven by the \ac{VSI} in thermally stratified disks, by creating synthetic images of molecular lines emitted from the model disks. For this purpose, we post-process the simulation data at the end of the runs ($t/t_\textrm{orb}=300$) using the Monte-Carlo radiative transfer code RADMC3D \citep{du2012}. We calculate the line-of-sight velocity maps from synthetic channel maps of $^{12}$CO, $^{13}$CO, and C$^{18}$O $J=2-1$ transitions with central frequencies at approximately 230.538 GHz, 220.399 GHz, and 219.560 GHz, respectively.

We assume a molecular abundance of $^{12}$CO to be $1.0\times10^{-4}$ relative to H$_2$ and calculate the abundances of $^{13}$CO and C$^{18}$O by adopting the isotope ratios $\rm [^{12}C]/[^{13}C]\sim77$ and $\rm [^{16}C]/[^{18}O]\sim560$ in the interstellar medium \citep{wr1994}. We then create the 3D data cubes for the CO number density and the three velocity components from the simulation data. Using the information for CO lines from the LAMDA database \citep{sv2005} we construct the molecular line emission images of the model disks. 

We consider three inclination angles $i = 5^\circ$, $20^\circ$, and $35^\circ$ of the disk relative to the line of sight, while fixing the position angle at $90^\circ$ and the distance to the disk at $100\unit{pc}$. The synthetic cubes have a total bandwidth of $6\unit{km\ s^{-1}}$ and 120 channels with resolution of $0.05\unit{km\ s^{-1}}$. To assess the observability of the \ac{VSI} signatures, we convolve the data with a circular Gaussian beam with a \ac{FWHM} of $0.1''$ and add root-mean-square noise of $\sim 1.5\unit{mJy\ beam^{-1}}$ using \textsc{syndisk}\footnote{\url{https://github.com/richteague/syndisk}}. 

After constructing the synthetic channel maps, we take the following steps to extract kinematic information of the observed \acp{PPD}. We derive the velocity centroid map from the synthetic images using \textsc{BETTERMOMENTS}\footnote{\url{https://github.com/richteague/bettermoments}} \citep{tf2018}. At each pixel, we fit the line profile using a Gauss-Hermite function to obtain the line-of-sight velocity distribution of the disk. Subsequently, we utilize \textsc{EDDY}\footnote{\url{https://github.com/richteague/eddy}} \citep{t2019} along with the Keplerian disk model (outlined below) to subtract it from the velocity centroid, $v_0$, thereby isolating the velocity perturbations.

We follow the methodology outlined by \citet{bf2021} to construct the Keplerian disk model, $v_{\text{mod}}$, as follows. We first project the Keplerian velocity onto the plane of the sky, given by
\begin{align}
  v_{\phi} &= v_{\phi,100}\left(\frac{R}{100\unit{au}}\right)^{q_{\phi}}, \\
  v_{R} &= v_{R,100}\left(\frac{R}{100\unit{au}}\right)^{q_{R}}, \\
  v_{\text{mod}} &= v_{\text{LSR}} + v_{\phi}\cos\varphi\sin i + v_R\sin\varphi\sin i, \label{eq:vmod}
\end{align}
where $v_{\phi,100}$ and $v_{R,100}$ are the azimuthal and radial velocity of the disk at 100 au from the central star, respectively, with the power-law slope of $q_\phi$ and $q_R$. In \cref{eq:vmod}, $v_{\text{LSR}}$ represents the systematic velocity of the disk and $\varphi$ is the polar angle measured east of north relative to the red-shifted major axis. Additionally, we account for the effect of the elevated emission surface by adopting the approach of \citet{lt2021}, who define the CO emission surface as
\begin{equation}
  \tilde{z}(\tilde{R}) = \tilde{z}_0 \left(\frac{\tilde R}{1''}\right)^{\Psi}\exp\left(-\left[\frac{ \tilde R}{\tilde{ R}_{\text{taper}}}\right]^{q_{\text{taper}}}\right),
\end{equation}
where $\tilde{z}_0=\tilde{z}(\tilde{R}=1'')$ and $\Psi$ describes the flaring of the emission surface, and $\tilde{R}_{\text{taper}}$ and $q_{\text{taper}}$ parameterize the tapering of the emission surface at the disk edges. Here, all quantities denoted with a tilde are expressed in units of arcseconds. 

\begin{figure*}
  \plotone{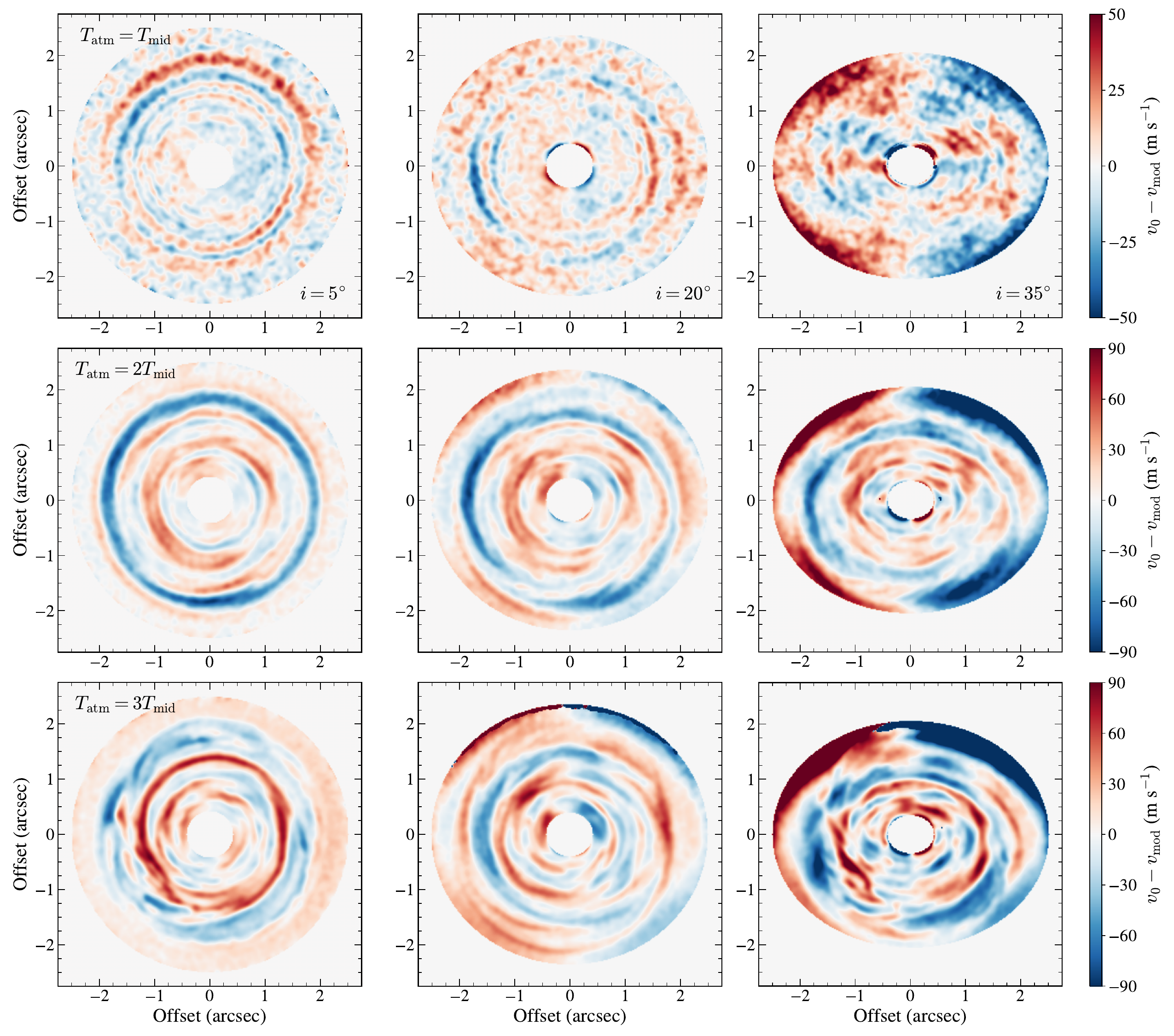}
  \caption{Distributions of the residual velocity, $v_0-v_\textrm{mod}$, of synthetic $^{12}$CO $J=2-1$ lines, observed by a circular beam with \ac{FWHM} of $0.10''$ after adding a root-mean-square noise of $\sim 1.5\unit{mJy}$. Each row corresponds to the disk with $n=1,2,3$ from left to right, while each column is for the inclination angle of $i=5^\circ, 20^\circ, 35^\circ$ from top to bottom.}
  \label{fig:vres}
\end{figure*}

\begin{figure}
    \plotone{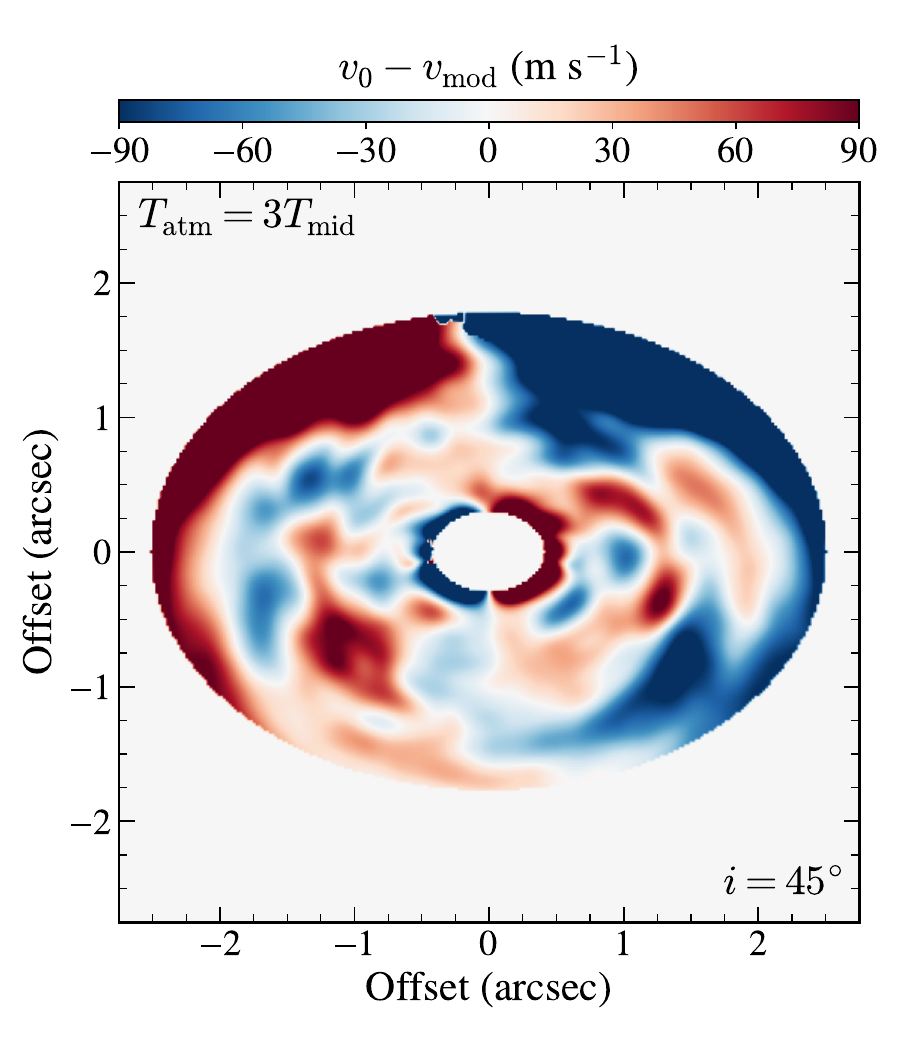}
    \caption{Velocity residual map for the $n=3$ disk viewed at an inclination of $i=45^\circ$. Several ring segments generated by \ac{VSI} perturbations are visible in the molecular line emission image.}
    \label{fig:i45}
\end{figure}

Employing a series of Markov Chain Monte Carlo models consisting of 128 walkers and 2000 burn-in steps, we identify the Keplerian disk model that best aligns with the obtained velocity centroid $v_0$. We iterate over a set of 13 parameters: the position of disk center, disk position angle, velocity profile parameters $(v_{\text{LSR}}, v_{\phi,100}, v_{R,100}, q_\phi, q_R)$, and emission surface parameters $(\tilde{z}_0, \Psi, \varphi, \tilde{R}_{\text{taper}}, q_{\text{taper}})$. We keep the distance to the source at $100\unit{pc}$ and adopt the same disk inclination angle as in the synthetic image. To mitigate potential contamination from emissions originating from the far side of the disk, we confine the radial range to $[0.55'', 2.0'']$, $[0.58'', 1.85'']$, and $[0.6'', 1.7'']$ for $i=5^\circ$, $20^\circ$, and $35^\circ$, respectively.
Additionally, we apply a mask to the velocity residual to remove the outer regions of the image that are added only during the synthetic observation process and are irrelevant to the actual motion of the disk.

\Cref{fig:vres} plots the observed maps of the velocity residual, $v_0-v_\textrm{mod}$, of synthetic $^{12}$CO $J=2-1$ molecular lines for differing inclinations $i=5^\circ$, $20^\circ$, and $35^\circ$ from left to right. Each row corresponds to the disk with $n=1$, 2, and 3 from top to bottom. The emergence and nonlinear saturation of the \ac{VSI} induce deviations in the velocity fields from the Keplerian rotation. These velocity deviations are more pronounced in a disk with stronger thermal stratification. For instance, at an inclination angle of $i=20^\circ$, the maximum amplitudes of residual velocities reach approximately 50, 80, and $100\unit{m}\;\unit{s}^{-1}$ in disks with $n=1$, 2, and $3$, respectively.

The shape of \ac{VSI} perturbations in velocity residual maps is crucial for identifying the presence of the \ac{VSI} in kinematic observations \citep{pt2023}. In the isothermal disk, velocity deviations from \ac{VSI} perturbations appear as quasi-axisymmetric rings at low inclinations ($i\leq 20^\circ$). The velocity residual map at $i=35^\circ$, shown in the top-right panel of Figure \ref{fig:vres}, displays axisymmetric rings alongside spoke-like radial features. These radial structures are most likely due to discrepancies between the true and fitted emission surfaces \citep{bf2021}. Notably, recovering the emission surface and rotation velocity becomes increasingly challenging as the inclination rises.

In thermally stratified disks, 
the \ac{VSI}-induced perturbations in velocity residual maps exhibit greater complexity compared to their isothermal counterparts. \Cref{fig:vres} shows that in such disks, velocity perturbations appear as ring segments even at low disk inclinations. This behavior arises from the non-monotonic distribution of the vertical shear $q$, which suppresses the development of coherent structures, as shown in \Cref{fig:per1,fig:per2}. Vortices generated by the Kelvin-Helomholtz-like parasitic instability further contribute to breaking quasi-axisymmetric rings into segments. Although the complex shapes of \ac{VSI} perturbations in thermally stratified disks can make detection challenging, the strong associated velocity deviations allow them to remain visible at inclinations as high as $i = 45^\circ$. This contrasts with the isothermal disk, where \ac{VSI}-induced perturbations are dominated by the vertical component, whose contribution to the line-of-sight velocity decreases at high inclinations (see \autoref{subsec:component} for discussion). For instance, Figure \ref{fig:i45} shows the velocity residual map for the $n=3$ disk at $i = 45^\circ$, revealing several ring segments.

\begin{figure*}
  \plotone{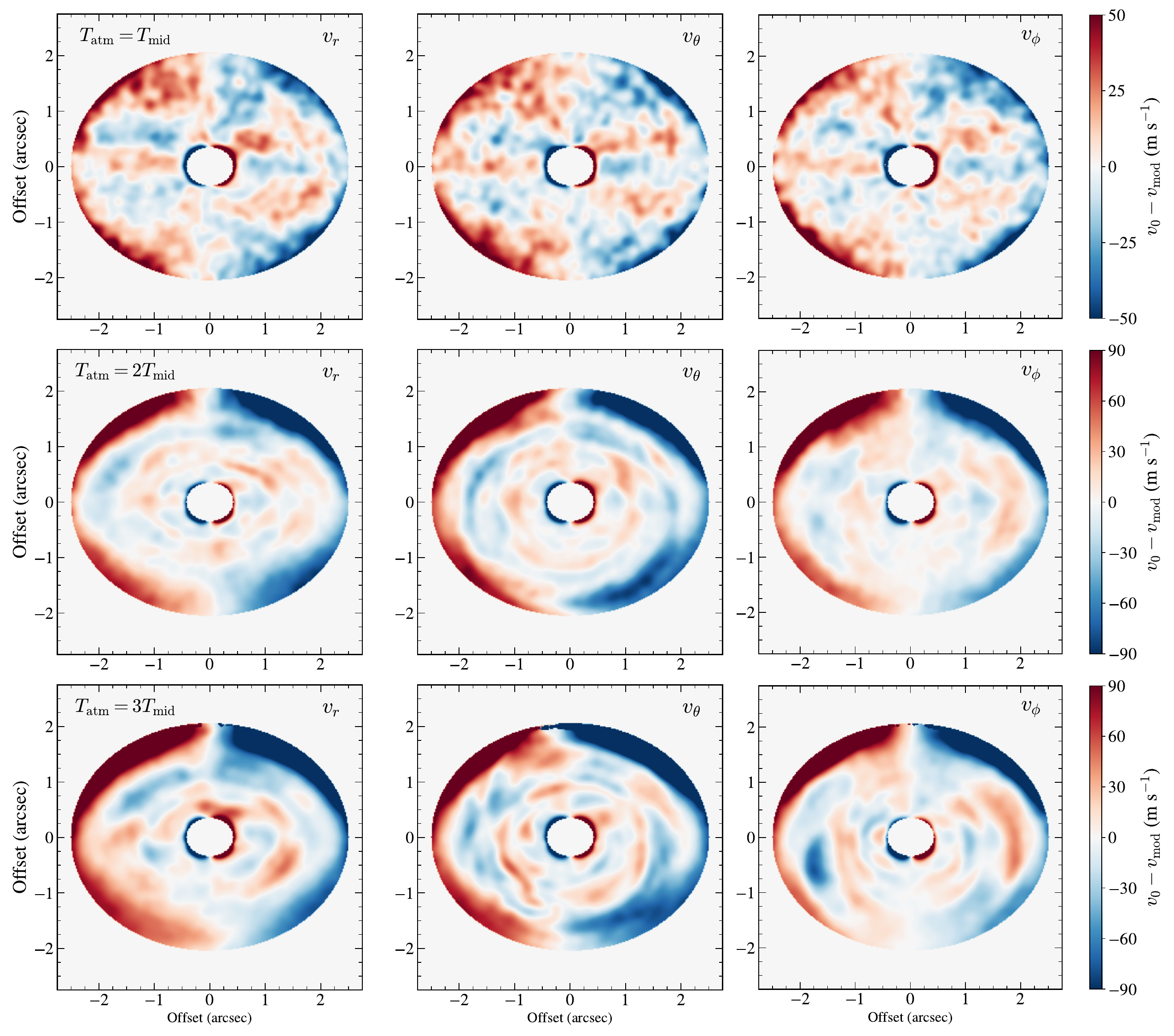}
  \caption{Contributions of (left) $v_r$, (middle) $v_\theta$, and (right) $v_\phi$ to the velocity residual, $v_0-v_\textrm{mod}$, for an inclination of $i=35^\circ$. Each row corresponds to the disk with $n=1,2,3$ from top to bottom. The contributions of all three velocity components are comparable at this inclination.}
  \label{fig:indv}
\end{figure*}

\section{Discussion} \label{sec:discussion}

We here discuss the contribution of each velocity component to the line-of-sight velocity and examine the effect of the optical depth of a tracer element through synthetic observations from different CO isotopologues. We also remark a few caveats that need to be considered in future studies.

\subsection{Contribution of Each Velocity Component} \label{subsec:component}

\citet{bf2021} demonstrated that the velocity deviations in their velocity residual maps for vertically isothermal disks are primarily due to the meridional velocity. As mentioned in \autoref{sec:HDsim}, our thermally stratified disks have almost comparable velocities in the radial and vertical directions, and about half smaller velocities in the azimuthal direction. To quantify the contribution of each velocity component to the total deviations, we create the data cubes where only one velocity component is taken from the simulations, while the other two components are set equal to the initial conditions (i.e., without perturbations). 

\Cref{fig:indv} plots the resulting velocity residual maps at an inclination of $i=35^\circ$. From left to right, each panel shows the maps generated when only $v_r$, $v_\phi$, or $v_\phi$ is extracted from the simulations, respectively. Each row is for the model with $n=1$, 2, and 3 from top to bottom. The maximum values of $|v_0-v_\text{mod}|$ are $\sim$ 35, 35, and 34$\unit{m\; s^{-1}}$ when the perturbed component is $v_r$, $v_\theta$, and $v_\phi$, respectively, in the isothermal disk. These values changes to $\sim$ 47, 48, and 31$\unit{m\ s^{-1}}$ in the $n=2$ disk, and $\sim$ 63, 55, and 58$\unit{m\ s^{-1}}$ in the $n=3$ disk, At $i=35^\circ$, all three components of velocity perturbations are comparable in magnitude.   At higher inclinations, however, the relative contributions of $v_r$ and $v_\phi$ become increasingly dominant compared to $v_\theta$.

\subsection{Effect of the Optical Depth}\label{subsec:optdepth}
\begin{figure}
  \plotone{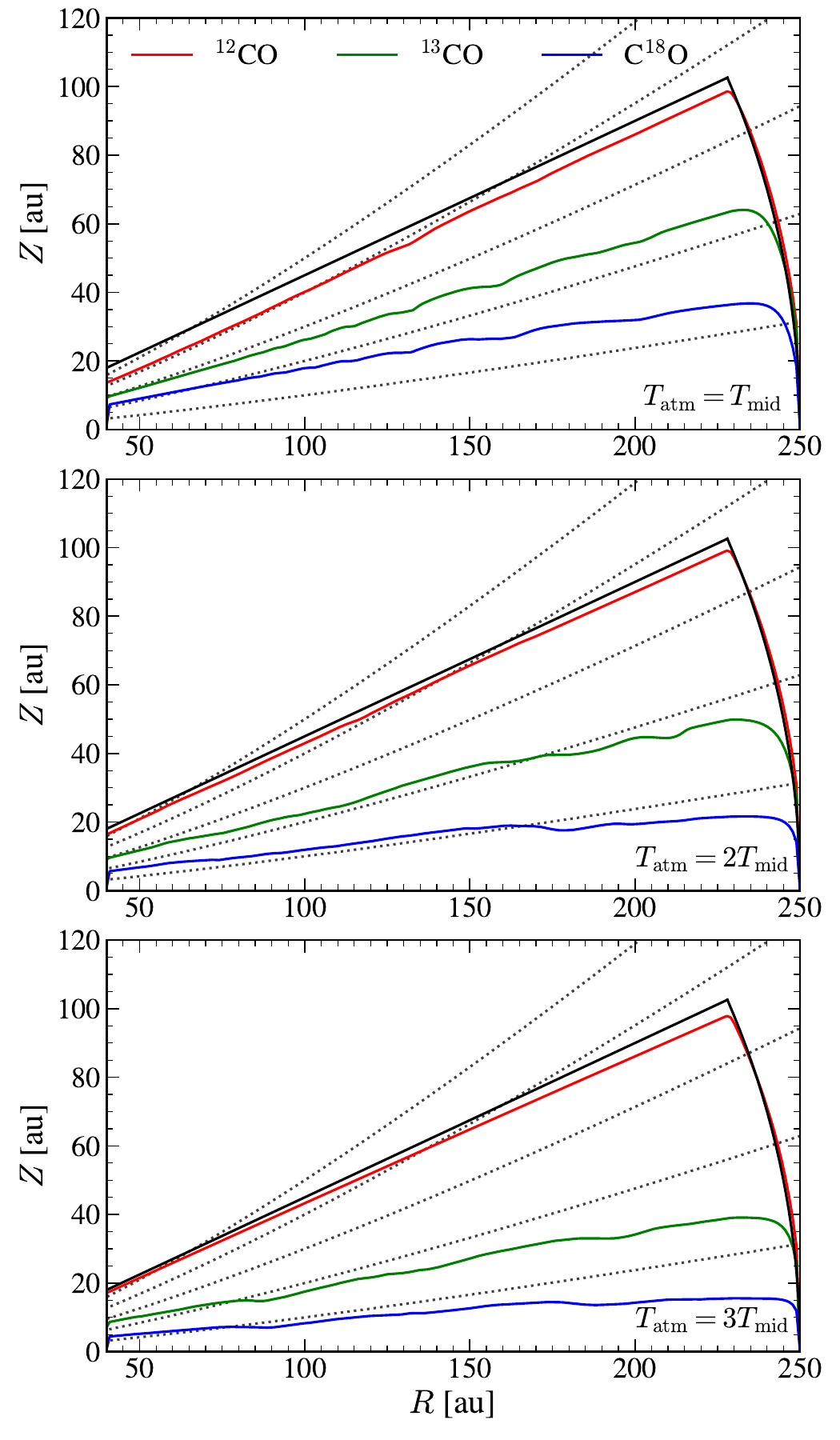}
  \caption{Locations of the emission surface with an optical depth of unity, observed along the $Z$-axis, for $^{12}$CO (red), $^{13}$CO (green), and C$^{18}$O (blue) lines in the disks with $n=1, 2, 3$ from top to bottom. 
  The solid black line denotes the simulation domain, while the dotted lines indicate the height with  $Z/H=1,\cdots,5$. }
  \label{fig:opt}
\end{figure}

For molecular line observations, probing the vertical disk structure is feasible using various isotopologues with different optical depths. \Cref{fig:opt} illustrates the heights at which the optical depth of the CO isotopologues reaches unity when observed along the $Z$-direction in the disks with $n=1, 2, 3$ from top to bottom, respectively. Given its higher abundance, $^{12}$CO predominantly traces regions near the disk surfaces, whereas the less abundant C$^{18}$O primarily probes regions closer to the midplane.\footnote{Physical quantities probed by $^{12}$CO may be subject to the adopted boundary conditions.} In addition, $^{13}$CO and C$^{18}$O are effective in tracing deeper regions within a more thermally stratified disk due to the steeper density profile in the vertical direction (see Figure 1 of \citetalias{yun2024}).

\begin{figure*}
  \plotone{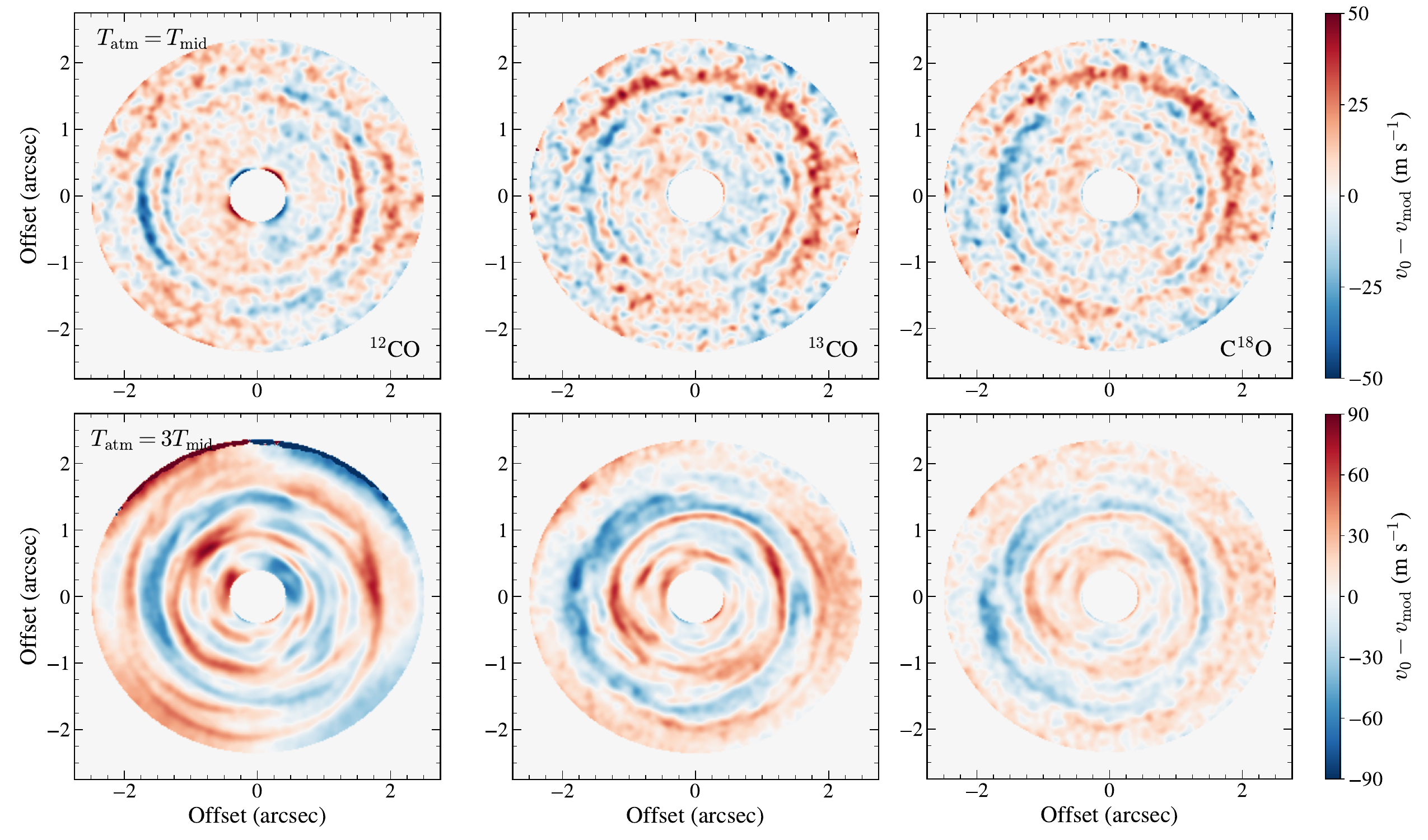}
  \caption{Velocity residual maps for
  $^{12}$CO, $^{13}$CO, and C$^{18}$O from left to right in the disks with $n= 1$ (top) and 3 (bottom) for an inclination of $i=20^\circ$. In the isothermal disk, the magnitudes of the velocity residuals remain small, irrespective of the tracer elements. In the $n=3$ disk, however, the velocity residual decreases as the tracers probe the regions closer to the midplane.}
  \label{fig:iso}
\end{figure*}

\Cref{fig:iso} compares the velocity residual maps for $^{12}$CO, $^{13}$CO, and C$^{18}$O observations from left to right. The top and bottom panels correspond to the disks with $n=1$ and 3, respectively, utilizing the same set of parameters as in \Cref{fig:vres}. It is apparent that for the isothermal disk, the effect of the optical depth on the velocity residual is not significant, consistent with the result of \citet{bf2021}. This is because the meridional velocity perturbation by the VSI in vertically isothermal disks does not exhibit strong height dependence (see the top panels of \cref{fig:per3}). 
In the thermally stratified disks, however, the strength of \ac{VSI}-induced features noticeably decreases going from $^{12}$CO to C$^{18}$O. This is attributed to C$^{18}$O tracing regions close to the midplane, while the regions with high-$|Z|$ experience strong shear and thus are more susceptible to \ac{VSI} in thermally stratified disks.

While the kinematic features of MWC 480 presented in \citet{tb2021} resemble the quasi-axisymmetric rings generated by \ac{VSI} perturbations, they cannot be fully explained by simulation results of \citet{bf2021} for a vertically isothermal disk. This discrepancy arises because the observed features exhibit a height dependence, with perturbations weakening closer to the midplane, and because they have a longer length scale than the synthetic images produced by vertically isothermal disk models. Thermal stratification, however, can resolve both issues by increasing the length scale of \ac{VSI} perturbations and introducing a height-dependent morphology. This suggests that the ring structures observed in MWC 480 may indeed originate from the \ac{VSI}.

\subsection{Comparison with Other Mechanisms}

In \autoref{sec:SynObs}, we have demonstrated that \ac{VSI}-driven velocity perturbations can be observed with \ac{ALMA} and that their strength is enhanced by thermal stratification. However, other mechanisms can also produce similar large-scale gas motions in \acp{PPD}. Notable mechanisms include planet-disk interactions \citep{da2020, pt2023} and vortices \citep{hi2018, rm2020}, although the resulting structures exhibit distinct kinematic features compared to those of the \ac{VSI}.

First, \ac{VSI}-induced features are global in the sense that they span large radial and azimuthal ranges. They appear as quasi-symmetric rings in vertically isothermal disks and ring segments in thermally stratified disks. Planet-induced spirals are also global but involve localized velocity kinks at the planet's position \citep{pd2015, pv2019}. Second, \ac{VSI}-induced perturbations in thermally stratified disks are strongest in high-$|Z|$ regions, whereas planet-induced spirals are more prominent in the midplane. Third, vortices generated, for example, by the Rossyby instability are local, appearing in a few channel maps, while \ac{VSI} perturbations are present in multiple channel maps with similar magnitudes. Therefore, distinguishing the \ac{VSI} signatures will require a closer examination of the velocity residual maps to find unique kinematic features of alternative scenarios.

\subsection{Caveats}

While our study presents the enhancement of the VSI perturbations due to thermal stratification, we note that our disk models are inviscid and isothermal, which is an ideal condition for the growth of the \ac{VSI}. Previous studies have shown that the inclusion of viscosity \citep{ng2013},  finite cooling time \citep{ly2015}, vertical buoyancy \citep{ly2015, lp2018}, and magnetic fields \citep{lp2018}
can stabilize the \ac{VSI} and weaken the resulting velocity perturbations. Recent numerical studies with vertically varying cooling time \citep{fo2023, pb2023} and radiation hydrodynamic simulations of irradiated disks \citep{mf2024, zz2024} commonly found the \ac{VSI} is damped in the optically-thick, cold regions where the cooling time is longer than the critical cooling 
time.  So, we expect that the amplitudes of \ac{VSI} perturbations presented in this study is likely to be upper limits compared to the ones in real disks.

In addition, various non-axisymmetric features produced by a massive embedded planet can interfere with \ac{VSI}-driven turbulence, making it difficult to distinguish between perturbations caused solely by the \ac{VSI} and those originating from the planet \citep[see also][]{zk2023, bf2024}. Depending on the planet mass, the planet-induced spiral wakes or a gap can disrupt or damp the \ac{VSI} perturbations, especially in the disk midplane \citep{bf2024}.

\section{Conclusion} \label{sec:conclusion}

In this paper, we investigate the effects of vertical thermal stratification on the nonlinear turbulent motions induced by the \ac{VSI} in \acp{PPD}. We consider three disk models in hydrostatic equilibrium, characterized by $n=T_\textrm{atm}/T_\textrm{mid}=1,2,3$:  $n=1$ represents an isothermal disk, while $n=2$ and $3$ correspond to thermally stratified disks. We study their nonlinear evolution by conducting \ac{3D} hydrodynamic simulations. Once the disks reach a turbulent state at saturation, we generate various synthetic images of CO isotopologues lines with different disk inclinations. By extracting the velocity perturbations from these images, we assess the observability of \ac{VSI}-induced features using ALMA. Our main results can be summarized as follows:

\begin{itemize}
  \item[1.] 
 The \ac{VSI} perturbations exhibit stronger growth in a more thermally stratified disk, resulting in a higher level of turbulence at saturation. In terms of the Reynolds stress responsible for angular momentum transport to thermal pressure, $\alpha_{R\phi}=1\times 10^{-4}$ in the isothermal disk, which increases to $1\times 10^{-3}$ and $2\times 10^{-3}$ in the thermally stratified disks with $n=2$ and $3$, respectively.
  
  \item[2.] Synthetic observations of the turbulent disks induced by the \ac{VSI} reveal distinct residual velocities. In isothermal disks, these velocities appear as quasi-axisymmetric rings, while in thermally stratified disks, they manifest as ring segments, once the Keplerian rotational velocity is subtracted. At an inclination angle of $i=20^\circ$, the amplitude of the residual velocities amounts to $\sim 50$, 80, and $100\unit{m} \;\unit{s}^{-1}$  in the disks with $n=1$, 2, and $3$, respectively. These levels of the \ac{VSI} signatures can potentially be observable by the \ac{ALMA} despite their complex shapes induced by the non-monotonic vertical shear profile resulting from the thermal stratification as well as Kelvin-Helmholtz-like parasitic instability.
  
  \item[3.]  The enhancement of the radial and the azimuthal components of the \ac{VSI}-driven velocity perturbations with increasing thermal stratification of the disk enables kinematic features, in the form of ring segments, to remain observable even at disk inclinations as high as $i= 45^{\circ}$. The magnitude of the observed velocity perturbations depends on the tracer used. While the effect of the optical depth is insignificant in the isothermal disk, in thermally stratified disks, the velocity perturbations decrease remarkably when transitioning from $^{12}$CO to $^{13}$CO to C$^{18}$O, as the latter probes closer to the midplane where vertical shear is weak. 
\end{itemize}

\section*{Acknowledgments}

We are grateful to the referee for detailed and 
constructive comments.
The work of H.-G.Y.\ was supported by the grants of Basic Science Research Program through the National Research Foundation of Korea (NRF) funded by the Ministry of Education (2022R1A6A3A13072598). The work of W.-T.K.\ was supported by a grant of the National Research Foundation of Korea (2022R1A2C1004810). JB acknowledges support from NASA XRP grant No. 80NSSC23K1312.

\bibliographystyle{aasjournal}
\bibliography{vsipaper}

\appendix

\section{Test Simulations}\label{sec:test}

In the appendix, we demonstrate the reliability of our numerical results by examining the effects of numerical resolution and boundary conditions on the models with $n=1, 2$m and 3. To assess convergence, we employ the ratio of volume-averaged Reynolds stress to thermal pressure as defined in \cref{eq:alpha}.

\subsection{Resolution Test} \label{subsec:restest}

\begin{figure}[h]
  \epsscale{0.6}
  \plotone{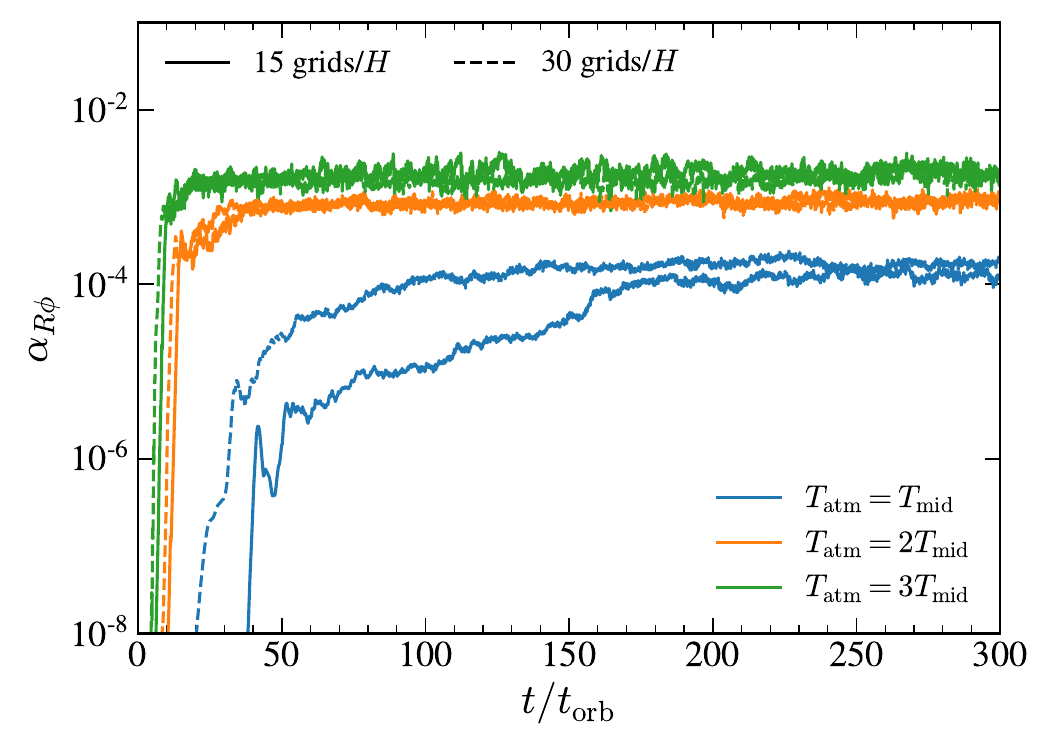}
  \caption{Comparison of $\alpha_{R\phi}$ between models with different resolutions. The solid and dashed lines represent the models with 15 and 30 cells per scale height, respectively. While $\alpha_{R\phi}$ varies with resolution at $t/t_\textrm{orb}\lesssim 250$ in the vertically isothermal disk, it shows almost no dependence on resolution in the thermally stratified disk. In both cases, the saturated value of $\alpha_{R\phi}$ remains essentially unaffected by resolution.}
  \label{fig:restest}
\end{figure}

\citet{ff2020} recently argued accurately capturing the \ac{VSI} in their 2.5D simulations requires resolving disks with more than 64 cells per scale height. Our 3D disk models presented in the main sections adopt a resolution of 15 cells per scale height, lower than the minimum suggested resolution by \citet{ff2020}. To assess if our results are affected by this resolution, we conduct additional 3D simulations for models with $n=1, 2$, and 3, resolved by 30 cells per scale height. 

\cref{fig:restest} compares the values of $\alpha_{R\phi}$ from the high-resolution runs with those from the standard runs. We find the temporal changes in $\alpha_{R\phi}$ depend on the resolution at $t/t_\textrm{orb}\lesssim 200$ in the $n=1$ model. However, the difference is almost negligible in $n=2$ and 3 models. Also, saturated values of $\alpha_{R\phi}$ are insensitive to resolution in all models. This confirms that our models with 15 cells per scale height produce reliable results in terms of saturated turbulence levels and the presence of quasi-axisymmetric structures in synthetic maps.

\subsection{Boundary Condition Test} \label{subsec:bctest}

\begin{figure}[h]
  \epsscale{0.6}
  \plotone{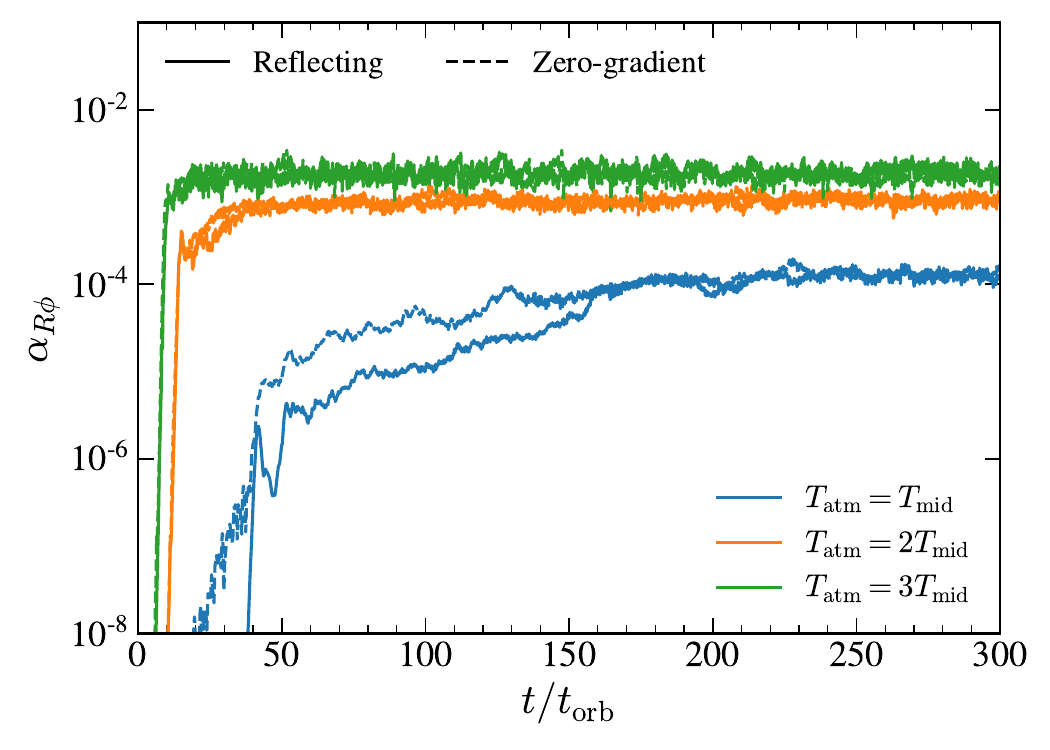}
  \caption{Comparison of $\alpha_{R\phi}$ from models with reflecting ($\textit{solid}$) and zero-gradient ($\textit{dashed}$) boundary conditions. Irrespective of vertical thermal stratification, saturated turbulence shows little sensitivity to the meridional boundary conditions at $t/t_\textrm{orb}\gtrsim200$.}
  \label{fig:bound}
\end{figure}

One characteristic feature of the \ac{VSI} is the formation of vertically elongated structures \citep{ng2013, lp2018}. As these structures typically develop near the disk surface in isothermal models, the boundary conditions at the meridional boundaries may potentially alter the outcomes of hydrodynamic simulations. The models outlined in the main sections utilize reflecting boundary conditions at the meridional boundaries. Here, we compare the results using reflecting and zero-gradient boundary conditions, the latter allowing gas to flow both inwards and outwards.

\cref{fig:bound} plots the resulting $\alpha_{R\phi}$ for the models with $n=1,2$ $3$. In the isothermal disk, $\alpha_{R\phi}$ depend on the boundary conditions at $t/t_\textrm{orb}\lesssim 200$, eventually converging to a constant value $\alpha_{R\phi}\sim10^{-4}$ at later times. 
In the thermally stratified disk, the effects of boundary conditions on turbulence levels are nearly negligible at almost all times. This suggests that the turbulence induced by the \ac{VSI} remains largely unaffected by the choice of boundary conditions. This validates our findings regarding the observability of the \ac{VSI} in thermally stratified disks.

\end{document}